\documentclass[12pt]{iopart}

\newcommand{\eq}[1]{\begin{equation}#1\end{equation}}

\newcommand{\ee}{\mathrm{e}}
\usepackage{graphics}
\usepackage{graphicx}

\begin{document}

\title{Evolution of entanglement after a local quench}
\author{Viktor Eisler and Ingo Peschel}
\address{Fachbereich Physik, Freie Universit\"at Berlin, Arnimallee 14,
D-14195 Berlin, Germany}
\eads{\mailto{eisler@physik.fu-berlin.de} and \mailto{peschel@physik.fu-berlin.de}}

\begin{abstract}

We study free electrons on an infinite half-filled chain, starting 
in the ground state with a bond defect. We find a logarithmic increase
of the entanglement entropy after the defect is removed, followed by
a slow relaxation towards the value of the homogeneous chain. The
coefficients depend continuously on the defect strength.

\end{abstract}

\section{Introduction}

The entanglement properties of quantum chains in their ground state
are by now rather well known. The most common measure is the
entanglement entropy calculated with the reduced density matrix for a
subsystem, usually a segment of the chain. This quantity was first
studied within field theory \cite{Wilczek} and more recently also for
various lattice models \cite{Vidal,Korepin,Keating,CC04,PeschelXY,Eisler}.
For non-critical systems it is finite
and of order one, but for critical systems it diverges logarithmically
with the length of the subsystem. Moreover, the prefactor of the
logarithm is given by the central charge $c$ in the conformal 
classification, see \cite{CC04}. In systems with defects or disorder
it may be modified \cite{Peschel05,Refael,Laflo}.
\par
A more complex situation arises if the quantum state evolves in time.
The simplest way to achieve this is a quench, where one changes a
parameter of the system instantaneously. An eigenstate 
of the initial Hamiltonian then becomes a superposition of the 
eigenstates of the final one and a complicated dynamics results.
The concomitant evolution of the entanglement entropy has
been the topic of several recent studies \cite{CC05,Dechiara06,
Eisert/Osborne06,Bravyi06}. If one is dealing with critical models,
this can be discussed within conformal field theory \cite{CC05}.
\par
So far, most quenches considered were global, i.e. the system
was modified everywhere in the same way. Such a quench can actually 
be realized for atoms in optical lattices \cite{Greiner02}.
In the theoretical studies,
it was found that the entanglement entropy first increases linearly
in time and then saturates at a value proportional to the size of the
subsystem. Thus it becomes an extensive quantity, in contrast to the
equilibrium case. These features can be understood in a simple 
picture where pairs of particles transmit the entanglement from
the initial state to later times \cite{CC05,Dechiara06}.
\par
In the present paper we study a different situation, namely a local
quench. We consider free electrons hopping on a chain which initially 
contains a defect in form of a weakened bond. This defect is suddenly 
removed and the electronic system has to readjust. The set-up is similar 
to the X-ray absorption problem in metals where the creation of a deep 
hole leads to a local scattering potential. When a conduction electron 
fills the hole, this potential is switched off again \cite{Mahan}.
\par
In contrast to a related study of the transverse Ising model 
\cite{Skrov06}, we start from the ground state.
We consider a 
section of the chain containing the defect in the interior or at its 
boundary and calculate the time evolution of its entanglement entropy 
with the rest. For the equilibrium case, this problem has been studied 
before \cite{Peschel05}. We find a time dependence which is very different 
from that for a global quench. The entanglement entropy only changes after 
a certain waiting time, increases then to a maximum or plateau, 
depending on the location of the defect, and finally decreases very 
slowly and in a universal way towards its equilibrium value. In particular, 
it remains always non-extensive. 
\par
The calculations are numerical and based on the determination of the reduced 
density matrix $\rho$ from the one-particle correlation functions.
The necessary formulae are given in section \ref{sec:model}.
The case of a central defect, typical 
single-particle eigenvalue spectra and the long-time behaviour of S 
are discussed in section \ref{sec:center}. The evolution of the entropy
at intermediate times and for various defect positions is presented in section 
\ref{sec:defpos}. The case of a boundary defect is studied in section 
\ref{sec:boundary} and described by simple formulae based on fronts moving
through the system. In section \ref{sec:summary} we sum up our findings.
In the two appendices we discuss the initial entanglement and present
a simple example of a global quench for comparison.

\section{Model}
\label{sec:model}

We consider a system of free spinless fermions hopping between neighbouring
sites of an infinite chain (XX model). The system is half filled and initially
prepared in the ground state of the inhomogeneous Hamiltonian
\eq{H_0 =- \frac 1 2 \sum_{n=-\infty}^{\infty}  t_n 
(c_n^{\dagger} c_{n+1} + c_{n+1}^{\dagger} c_{n}) \, ,}
where the hopping amplitudes are $t_0=t'$ and $t_n=1$ for $n\ne 0$. Thus
one has a single bond defect between sites $0$ and $1$. At time $t=0$ this 
defect is removed and the time evolution is governed by the homogeneous
Hamiltonian
\eq{H_1 =- \frac 1 2 \sum_{n=-\infty}^{\infty}
(c_n^{\dagger} c_{n+1} + c_{n+1}^{\dagger} c_{n}) \, .}
\par
Both $H_0$ and $H_1$ describe critical systems.
Our aim is to study how the entanglement between a subsystem of length $L$ 
and the rest of the chain evolves after the quench.
The subsystem is chosen in such a way that the initial defect is inside
of it or at its boundary.
%
\begin{figure}[thb]
\center
\includegraphics[scale=0.6]{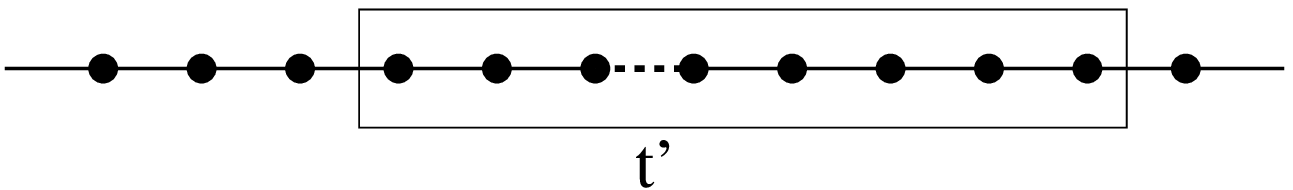}
\end{figure}
\par
The entanglement properties are determined by the reduced density matrix
which for free fermions has the following  diagonal form 
\cite{Cheong/Henley04,Peschel03}:
\eq{
\rho_L = \frac{1}{\tilde Z} e^{-\tilde H} \; , \quad
\tilde H = \sum_{k=1}^{L} \varepsilon_k(t) f_k^{\dagger} f_k \, .}
Here $\tilde Z$ is a normalization constant ensuring 
$\mathrm{tr}\,\rho_L = 1$ 
and the fermionic operators $f_k$ follow from the $c_n$ by an orthogonal
transformation. The eigenvalues $\varepsilon_k(t)$ are given by
\eq{
\varepsilon_k(t) = \ln \frac{1-\zeta_k(t)}{\zeta_k(t)},}
where $\zeta_k(t)$ are the eigenvalues of the time-dependent ($L \times L$)
correlation matrix
\eq{C_{jl}(t)=\langle 0| \, c_j^{\dagger}(t) \, c_l(t) \, |0 \rangle \, .}
Here $|0 \rangle$ is the ground state of $H_0$ and the indices $j$ and $l$ 
run over the sites of the subsystem. The $\zeta_k(t)$ also determine the 
entanglement entropy $S= -\mathrm{tr}(\rho_L \ln \rho_L)$ via
\eq{
S(t)=-\sum_k \zeta_k(t) \ln \zeta_k(t) - 
\sum_k (1-\zeta_k(t)) \ln (1-\zeta_k(t)) \, .
\label{eq:entropy}
}
\par
To obtain the matrix ${\bf{C}}(t)$ one diagonalizes the operator 
$H_1$ by a Fourier transform
\eq{H_1 = -\sum_q \cos q\; c^\dagger_q c_q \, .}
The time dependence of the $c_q$ then is
$c_q(t) = \exp (it\cos q)\, c_q$ and Fourier transforming back gives,
for a ring of $N$ sites,
\eq{c_j(t)=\sum_m U_{jm}(t) c_m \, , \quad 
U_{jm}(t)=\frac 1 N \sum_q \ee^{-iq(j-m)}\ee^{it \cos q} \, .}
In the thermodynamic limit, the matrix elements  $U_{jm}(t)$ of the unitary
evolution operator can be written as Bessel functions
and the correlation matrix becomes
\eq{C_{jl}(t)=i^{l-j} \sum_{m,n} i^{m-n} J_{j-m}(t) J_{l-n}(t) C_{mn}(0) \, .
\label{corellt}}
This expresses ${\bf{C}}(t)$ in terms of the matrix ${\bf{C}}(0)$
calculated with the initial state $|0 \rangle$.  The information 
is transmitted via the Bessel functions which have a maximum when 
the spatial separation is equal to the elapsed time. The factors 
in front of the Bessel functions can lead to imaginary correlations
and thus to local currents in the system. This is what one also 
expects on physical grounds.
\par
The initial matrix elements $C_{mn}(0)$ were already obtained in 
\cite{Peschel05} for the region to the right of the defect.
Then they have the form
\eq{C_{mn}(0)=C^0_{mn}-C^1_{mn} \, , \quad m,n > 0}
where $C^0_{mn}$ denotes the correlation matrix of the homogeneous
system
\eq{C^0_{mn}=\frac{\sin \left[ \frac \pi 2 (m-n)\right]}{\pi (m-n)}}
and $C^1_{mn}$ is the contribution of the defect, given explicitly 
in Eq. (8) of \cite{Peschel05}. It depends only on $m+n$ and vanishes
for $t'=1$. If the defect cuts the chain, $C^1$ reduces to $C^0$ 
with argument $m+n$.  When both sites are to the left of the defect,
one simply has to replace $m+n$ with $m+n-2$, and by a
straightforward generalization one can obtain corresponding formulae 
for $m$ and $n$ on opposite sides of the defect.
\par
With these initial values, the correlation matrix ${\bf{C}}(t)$ was
calculated numerically and then diagonalized to obtain the eigenvalue 
spectrum and the entanglement entropy. Since the Bessel functions decay 
rapidly for indices much larger than the argument $t$, one can 
confine each of the sums in (\ref{corellt}) to about $(2t+4L)$ terms. Most
calculations were done for times of the order 100-200.
\section{Central defect and spectra}
\label{sec:center}
We start with analyzing a simple symmetric situation, namely a defect
with $t'=0$ in the center of the subsystem. Thus one starts with two
uncoupled infinite half-chains which then are connected.
\par
\begin{figure}[thb]
\center
\includegraphics[scale=0.3,angle=270]{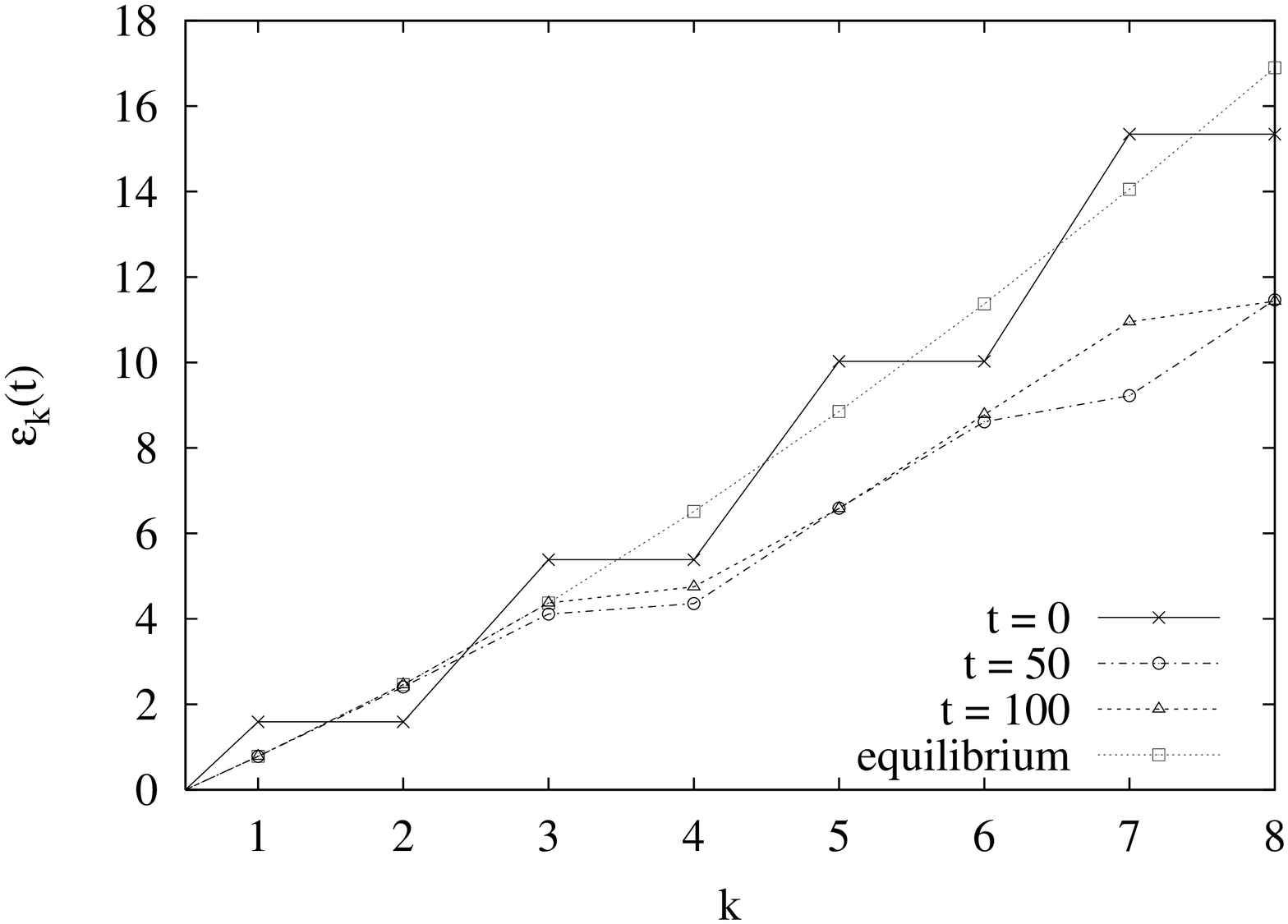}
\includegraphics[scale=0.3,angle=270]{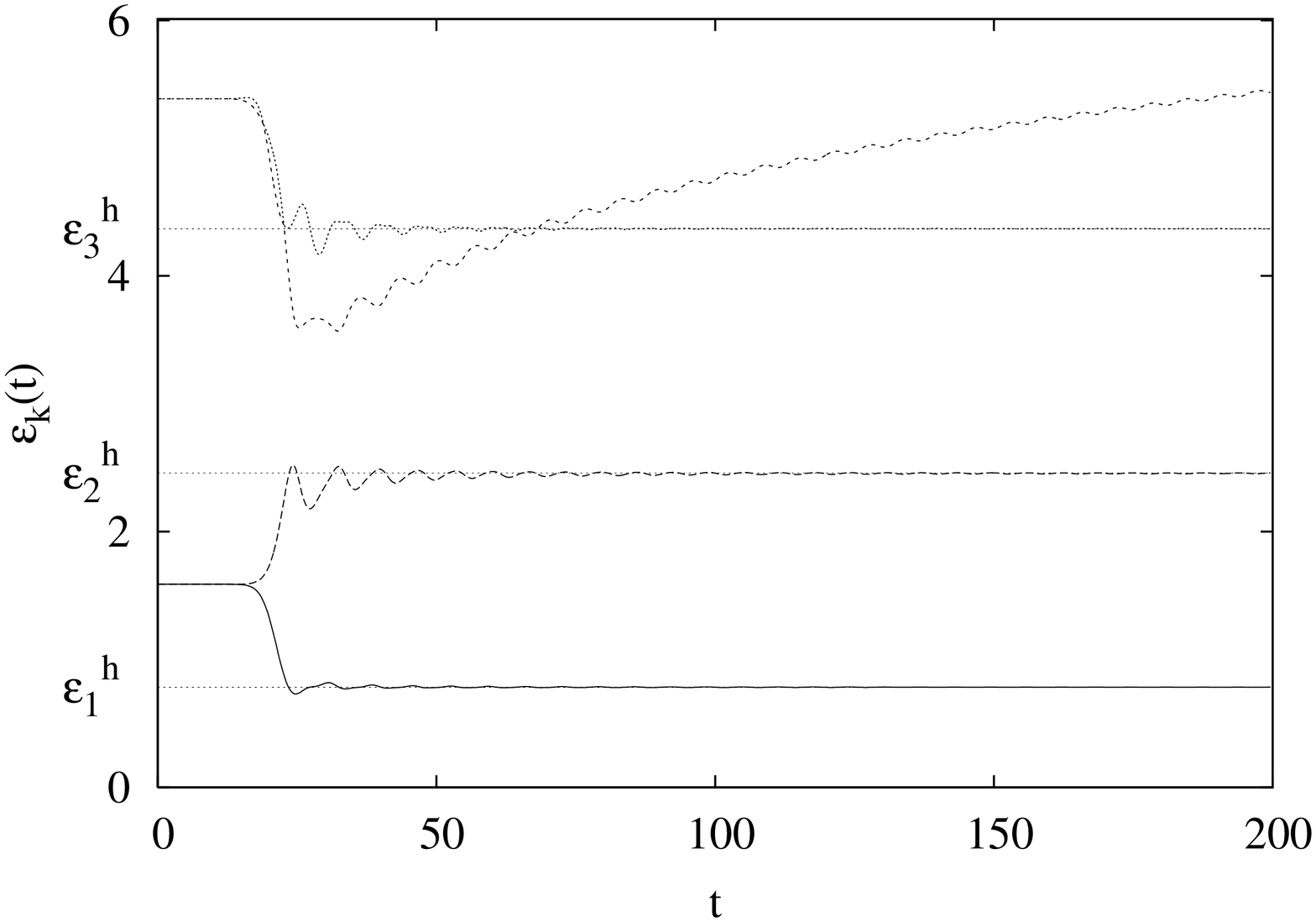}
\caption{Time evolution of the low-lying single-particle eigenvalues 
$\varepsilon_k(t)$ for a subsystem of $L=40$ sites with a central defect $t'=0$.
Left: snapshots of the positive eigenvalues for different
times, compared with the spectrum in equilibrium. Right: 
time evolution of the four lowest eigenvalues.}
\label{fig:epsilon_t}
\end{figure}

In Figure \ref{fig:epsilon_t} the time evolution of the single-particle
spectrum is shown in two different ways. On the left hand side,
the eight lowest eigenvalues $\varepsilon_k(t)$ are plotted in ascending 
order for several times. Only the positive ones are shown, since due to the
half-filling the spectrum is symmetric with respect to zero. These snapshots
show that the initial step structure resulting from the degeneracies of
the uncoupled chains smoothens with time and the dispersion curve seems to 
approach the one of the homogeneous system. However, two steps remain and
as a consequence the curves lie below the asymptotic one. Roughly speaking,
this corresponds to a stronger entanglement.
%
\par
On the right, the time evolution of the four lowest eigenvalues is displayed.
It shows two important features. Firstly, the eigenvalues remain unchanged
and the two-fold degeneracy survives up to a time $T\approx L/2$. 
This can be explained in terms of a front propagating with velocity $v=1$ 
\cite{Antal99} which is the Fermi velocity in the system and also the speed of
the maxima in the Bessel functions.
Before the front, starting at the defect, reaches 
the boundary, the subsystem to the right (left) of the defect cannot become
entangled with the environment to the left (right).
\par
Secondly, for $t>T$ three of the eigenvalues quickly
relax towards the values $\varepsilon_k^{h}$ of the homogeneous
system but one, on the contrary,
starts to evolve rather slowly. These ``anomalous'' eigenvalues
(another one is found among the higher levels) lead to the kinks in the
dispersion curves on the left of Fig. \ref{fig:epsilon_t} when they are close 
to another level. The smallest one is the most important for the entanglement,
and its time dependence will determine that of $S$, since the other 
$\varepsilon_k(t)$ are basically constant.  
\par
\begin{figure}[thb]
\center
\includegraphics[scale=0.4,angle=270]{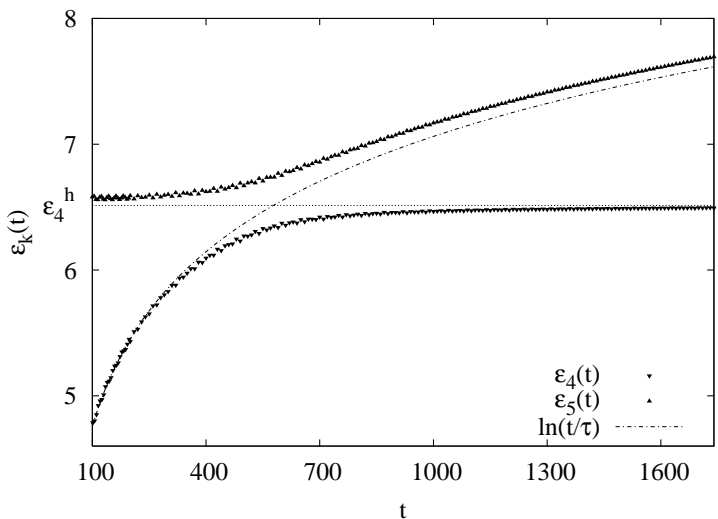}
\caption{Large-time behaviour of the eigenvalues $\varepsilon_4(t)$ and 
$\varepsilon_5(t)$ already seen in Figure \ref{fig:epsilon_t}.
The avoided level crossing can be fitted by $\ln (t /\tau)$
with $\tau \approx 1$.}
\label{fig:avoidcross}
\end{figure}

Therefore it is important to have a general picture of its behaviour.  
Figure \ref{fig:avoidcross} shows  what happens for times $t$ up to 1600.
In contrast to Figure \ref{fig:epsilon_t}, where the anomalous eigenvalue
simply crossed over, one sees an avoided crossing with the next one,
$\varepsilon_5$, which already looked relaxed at smaller times. This
can be attributed to the fact that the two eigenstates in this case
have the same reflection symmetry.
%
\par
At the avoided crossing, the two eigenvalues exchange role and
the anomalous parts of the curves can be well described by a single 
logarithm of the form $\ln (t /\tau)$, as shown in the Figure. Therefore 
the spectrum roughly looks as in equilibrium with at least one additional 
eigenvalue $\varepsilon_{an}$. One expects that for very large
times $\varepsilon_{an}$ finally converges to the maximal eigenvalue of 
the homogeneous system. Therefore the logarithmic behaviour cannot 
persist indefinitely. This could be confirmed for a rather small
subsystem with $L=6$ sites, but for larger $L$ the necessary times
are  numerically inaccessible. Moreover, the $\varepsilon_{k}$
can only be calculated reliably for values up to about 25.
\par
One can also look at the single-particle eigenfunctions connected
with the $\varepsilon_{k}$. Then one finds that they develop imaginary
parts which vanish again as
the eigenvalue approaches the equilibrium limit. In the anomalous eigenvector
they persist more and generally speaking this eigenvector looks more extended 
than the others when $\varepsilon_{an}$ is not close to a crossing.
\par
\begin{figure}[thb]
\center
\includegraphics[scale=0.4,angle=270]{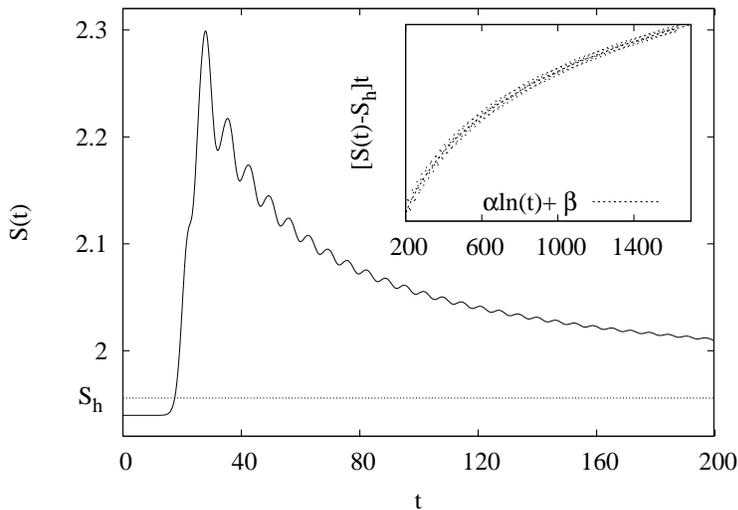}
\caption{Time evolution of the entanglement entropy for a subsystem of
$L=40$ sites with a central defect $t'=0$. A sudden jump is followed by a slow
relaxation towards the homogeneous value $S_{h}$. The inset shows the 
logarithmic correction to the $1/t$ decay.}
\label{fig:entropyevol}
\end{figure}

The entanglement entropy, finally, is shown in  Figure \ref{fig:entropyevol}.
The basic features are consequences of eigenvalue spectrum discussed above.
One has a sudden jump at $t= T$, followed by a very slow relaxation towards 
$S_{h}$, the value in the homogeneous system. For a large subsystem,
initial and final value of $S$ are also the same, as discussed in 
Appendix A.
%
\par
As mentioned above, the slow decay can be understood in terms of the anomalous 
eigenvalue $\varepsilon_{an}$. Its observed logarithmic time dependence 
implies $\zeta_{an}(t) \approx \tau / t$. This yields for the entropy
\eq{
S(t)=S_h+ \frac{\alpha \ln (t) + \beta}{t}.
\label{eq:entdecay}}
The inset in Figure \ref{fig:entropyevol} shows that the logarithmic corrections
to the $1/t$ time dependence can be indeed observed and fitted very well.
\par
The height of the maximum in $S$ depends on the size of the subsystem and one 
finds the behaviour
\eq{
S_m  -S_h = \frac{c_m }{3}\ln L + k_m \, ,
\label{eq:smax}
}
where $c_m \approx 0.23$ and $k_m \approx 0.06$.
\par
All these results were obtained for a defect with $t' = 0$. However, they
also hold for the general case $0 < t' < 1$. Then the spectra and the entropy 
have similar behaviour only the amplitude $c_m$  and the constant $k_m$ decrease
with increasing $t'$, since the effect must vanish for $t'=1$.
\section{Entropy and defect position}
\label{sec:defpos}
We now ask how the picture changes if one varies the position
of the defect. For simplicity we discuss the case $t'=0$,
in which the defect cuts the subsystem into two parts with $L_1$ sites 
to the left and $L_2=L-L_1$ to the right of it.
\par
\begin{figure}[thb]
\center
\includegraphics[scale=0.4,angle=270]{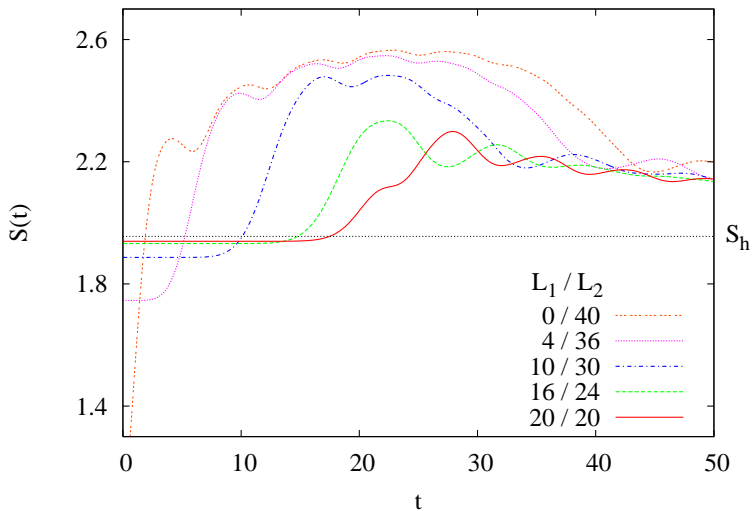}
\caption{Time evolution of the entropy for a subsystem of $L=40$
sites for several defect positions $L_1/L_2$,
indicating the number of sites to the left/right of the defect
with $t'=0$.
}
\label{fig:asymmblock}
\end{figure}

Figure \ref{fig:asymmblock} shows the entropy for intermediate times 
and several defect positions, specified by the numbers $L_1$ and
$L_2$. The extreme cases are a defect at the boundary, $L_1=0$,
and in the center  $L_1=L/2$, as in the previous section.
The main feature is the development of a plateau-like region,
which can be understood in terms of fronts propagating from the
defect site. The entropy increases rapidly at $T_1 \approx L_1$ 
when one of the fronts reaches the closest boundary 
and starts to decay at $T_2 \approx L_2$ when both fronts 
have left the subsystem. Thus the plateau is related to 
the asymmetry of the set-up.
%
\par
One also finds that at time $t=L_2$ where the plateau region ends,
the entropy always has the same value which moreover coincides with
the maximum value (\ref{eq:smax}) found for the central defect in 
the previous section. Beyond this point, the long-time region begins
and shows the same relaxation behaviour as in (\ref{eq:entdecay})
for all defect positions, although the curves do not coincide completely
when shifted appropriately. The plateau itself will be studied in more 
detail in the following section.
\section{Defect at the boundary}
\label{sec:boundary}

To obtain a better understanding of the development of the entropy plateau
we now investigate the situation when it is most pronounced. This is the case
for a defect located at the boundary of the subsystem. The behaviour 
for these intermediate times should be connected with the propagation
of a wavefront in the subsystem. That such a front actually exists, can be
seen from the single-particle eigenfunctions.
\par
\begin{figure}[thb]
\center
\includegraphics[scale=0.2,angle=270]{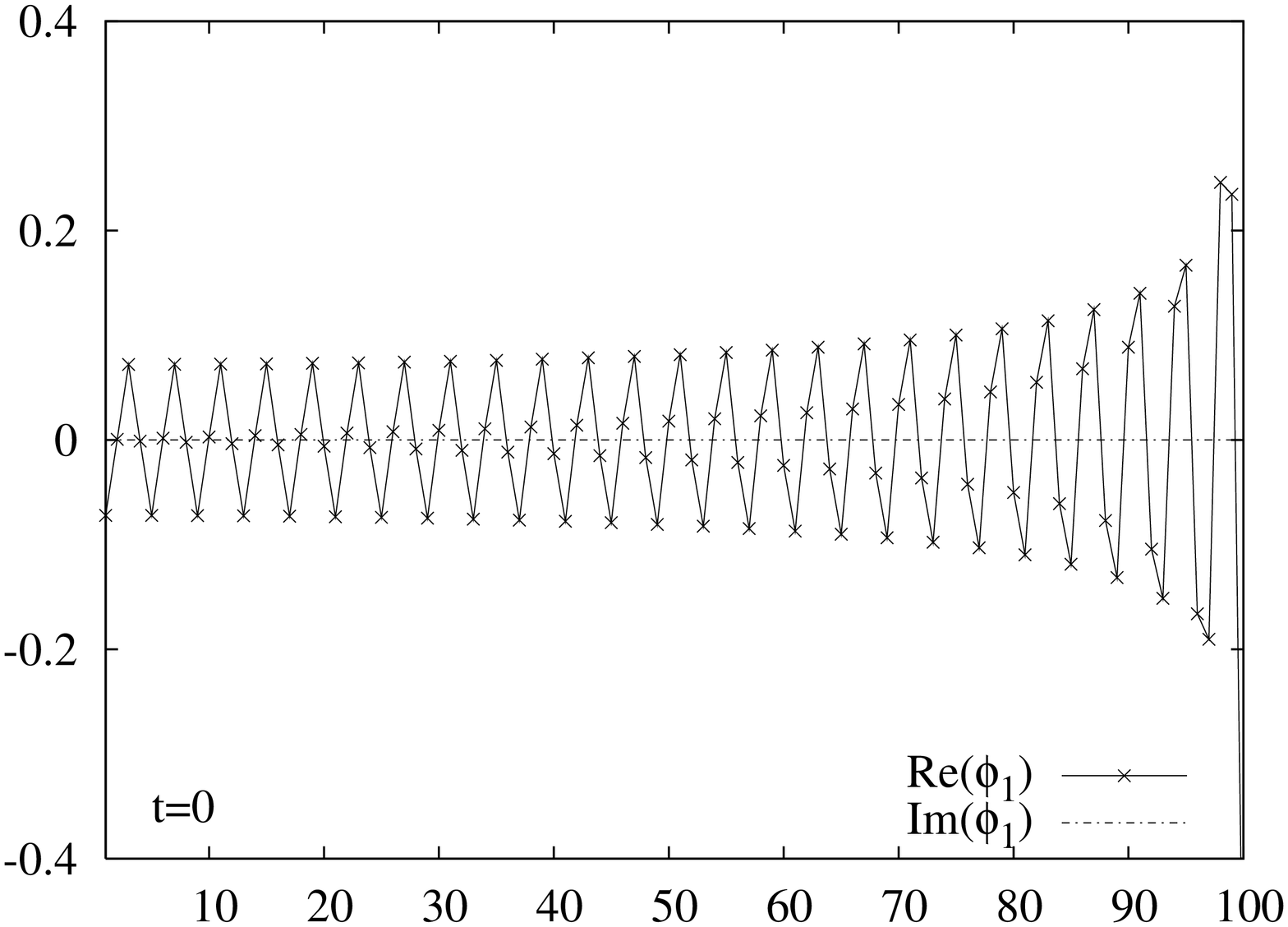}
\includegraphics[scale=0.2,angle=270]{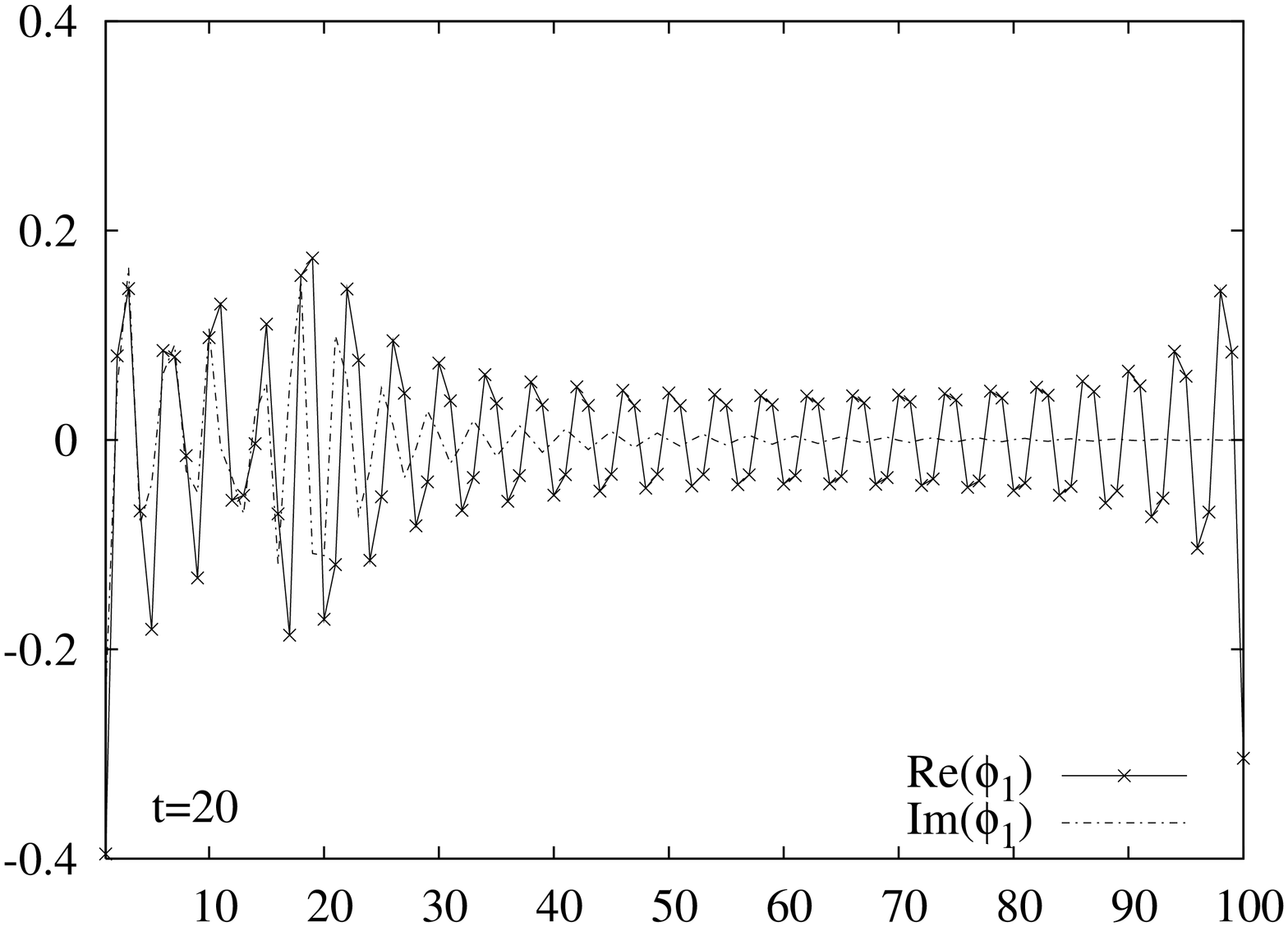}\\
\includegraphics[scale=0.2,angle=270]{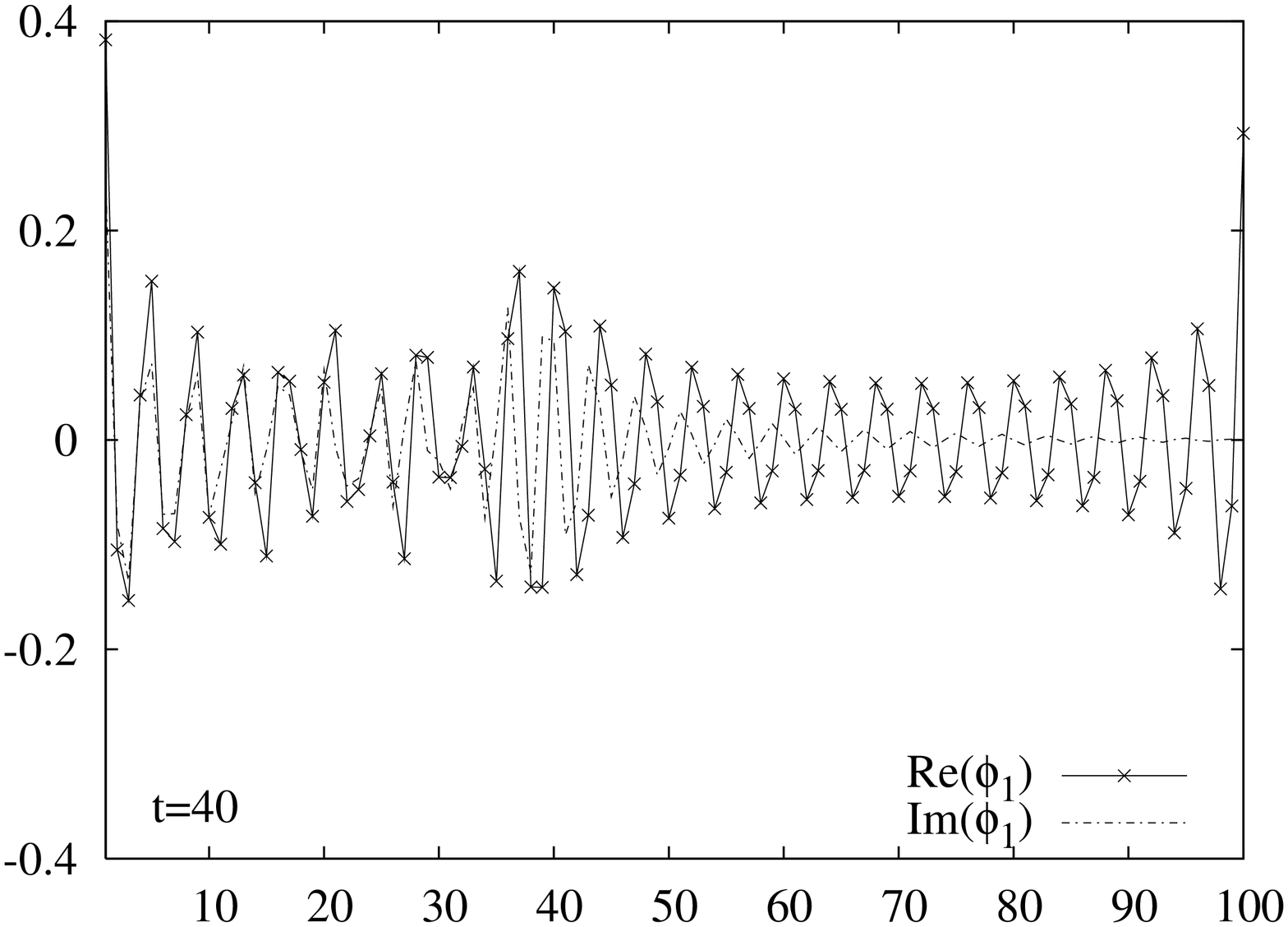}
\includegraphics[scale=0.2,angle=270]{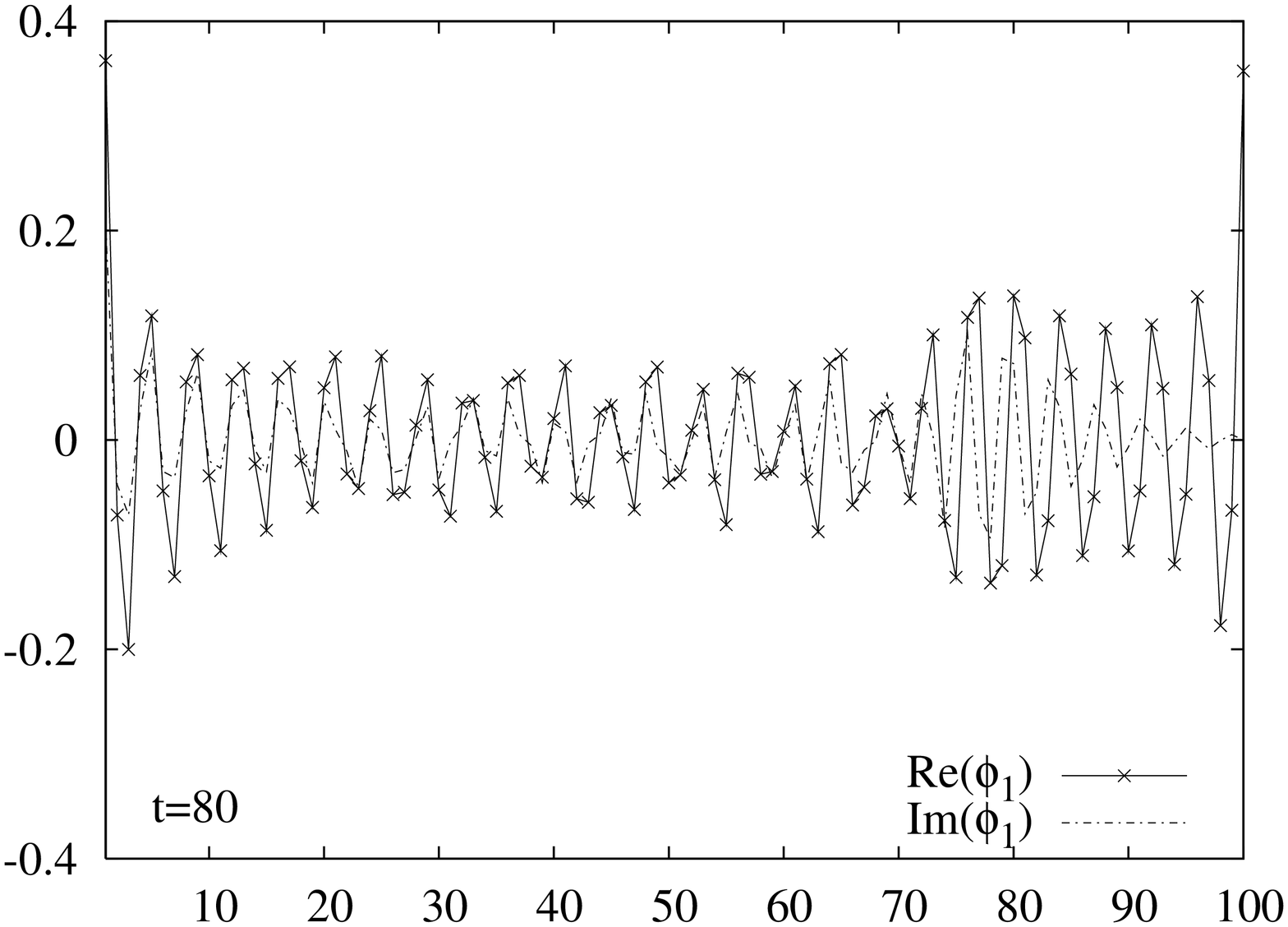}
\caption{Snapshots from the time evolution of the eigenvector corresponding
to the lowest lying single-particle eigenvalue of a subsystem with 
$L=100$ sites and with a defect $t'=0$ at the left boundary.}
\label{fig:eigvec}
\end{figure}

In Figure \ref{fig:eigvec} we show snapshots of the lowest eigenvector
for several intermediate times in a subsystem of $L=100$ sites.
One can clearly recognize the front, either from the real part or from
the imaginary part of the eigenvector. The latter, in particular, 
only develops behind the front, while it is approximately zero otherwise. 
%
\par
Furthermore, the form of the eigenvectors suggests to interpret the
front as an effective defect with a time-dependent position,
which divides the subsystem in two parts of size $t$ and $L-t$.
This leads to an analogy with the equilibrium problem where the defect is 
located inside a subsystem. Some details for this situation, which
was not covered in \cite{Peschel05} are given in Appendix A.
In that case, the entropy can be written as a sum of logarithmic terms,
which can be combined into a scaling function depending only on 
$x/L$ and $1-x/L$ where $x$ is the location of the defect.
\par
Starting from these equilibrium formulae, one is lead to make the following
generalized ansatz for the nonequilibrium case
\eq{
  f(t,L)=\frac{c_1}{3}\ln (1+t) + \frac{c_2}{3}\ln (1+L-t) + k,
\label{fitplateau}}
where each of the coefficients depends on the strength of the defect.
Note, that one has to have $c_1 \ne c_2$ in order to account for the 
nonsymmetric plateaus. The arguments of the logarithms
have been shifted to avoid divergences in the interval
$\left[ 0,L \right]$.
\par
In the simplest case of zero defect strength one can also motivate
the ansatz (\ref{fitplateau}) with the help of a simple physical picture.
Initially one has a subsystem which is entangled only with
one of the half-chains and corresponding entropy $S \sim 1/6 \, \ln L$. 
The propagating front carries information about the other half-chain, 
and therefore the $t$ sites of the subsystem which have already been 
visited by the front acquire an entropy $S \sim 1/3 \, \ln t$, while the 
other $L-t$ sites still remain entangled only with one half-chain. 
This gives the values $c_1=1$ and $c_2=1/2$ for the constants.
\par
\begin{figure}[thb]
\center
\includegraphics[scale=0.4,angle=270]{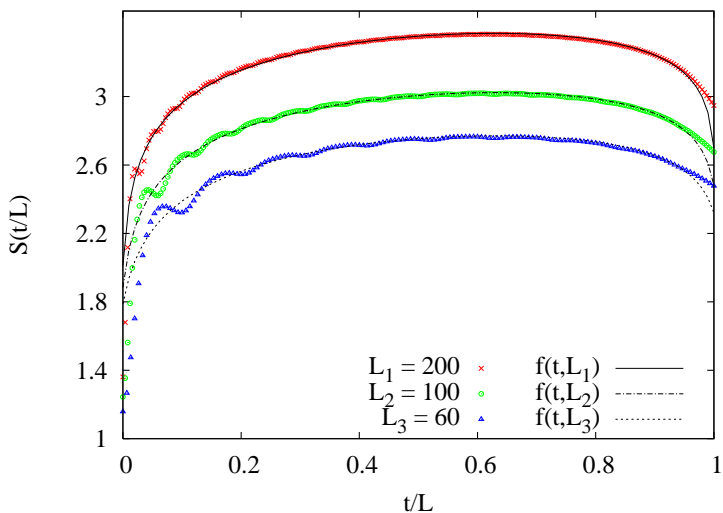}
\caption{Entropy plateaus in case of a $t'=0$ boundary defect 
for different subsystem sizes. The fitting function is defined
in Eq. (\ref{fitplateau}) and the coefficients were determined
by a single fit to the $L=200$ data.}
\label{fig:plateau}
\end{figure}

Figure \ref{fig:plateau} shows the entropy plateau curves with $t'=0$
for different system sizes, together with the fit functions (\ref{fitplateau}).
Our fit gave the values $c_1=0.96$, $c_2=0.55$ and $k=1.03$.
We emphasize, that these coefficients have been obtained via a single fit
to the $L=200$ data, and the other two curves differ only in the 
parameter $L$. One can see a good agreement with the data,
except for the boundaries of the time intervals. Indeed, for $t=L$
the fit formula (\ref{fitplateau}) scales as $S \approx 1/3 \, \ln L$,
that is the asymptotics of the equilibrium entropy $S_h$.
However, as it was mentioned in the previous section, the value $S(t=L)$
is described by (\ref{eq:smax}), and for later times one enters in the
regime of slow relaxation towards $S_h$.
%
\par
The situation becomes slightly more complicated for
arbitrary defect strengths in the range $0 \le t' \le 1$.
Considering the limiting case $t' = 1$ one has no time
dependence at all, thus $c_1=c_2=0$, however the entropy
in this homogeneous case must scale as $S \sim 1/3 \, \ln L$.
Therefore the constant in (\ref{fitplateau}) must be rewritten
to contain a term proportional to $\ln L$ with a new coefficient.
Finally, one could change to the scaling variables $t/L$ and
write the entropy in the form
\eq{
  S(t,L)=\frac{c_0}{3} \ln L +
  \frac{c_1}{3}\ln (t/L) + \frac{c_2}{3}\ln (1-t/L) + k',
\label{entscale}}
where all the parameters $c_0,c_1,c_2$ and $k'$ are functions of
the defect strength and have to be determined by fitting to the
data. Note, that the above form of the entropy is expected to hold
only for large $L$ and away from the boundaries of the time interval.
\par
\begin{figure}[thb]
\center
\includegraphics[scale=0.4,angle=270]{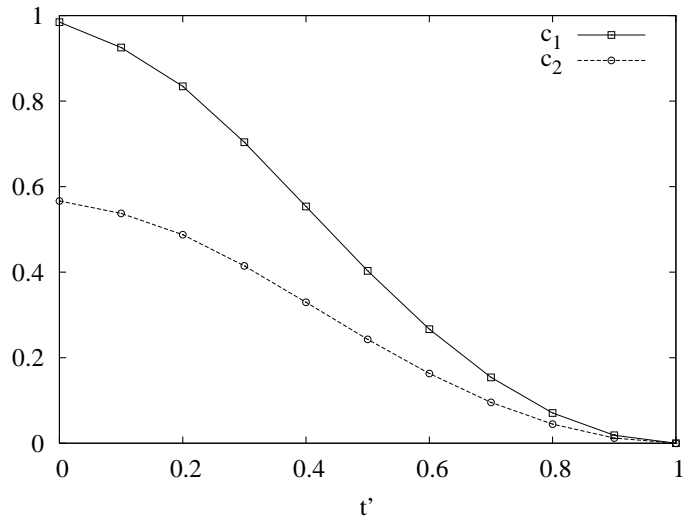}
\caption{Scaling function coefficients $c_1$ and $c_2$ as functions of the
defect strength.}
\label{fig:effc}
\end{figure}

In fact, we have always used the regularised ansatz (\ref{fitplateau})
to determine the $t'$-dependence of $c_1,c_2$ and $k$ by fitting to time series
with $L=100$ fixed. Then the additional logarithmic term was extracted from
fitting to the $k$ values obtained from time series with different $L$,
$t'$ being fixed. One finds after all, that $c_0 \approx 1+c_2$, thus
one needs only $c_1,c_2$ and $k'$ to describe the plateaus. The $t'$-dependence
of the former two is depicted in Figure \ref{fig:effc}. One can see that
the coefficients vary smoothly with $t'$, both of them going to zero
as $t' \to 1$.
%
\par
In conclusion, the time dependence of the entropy in the plateau-region
can be written in the scaling form (\ref{entscale}) for all values of the defect
strength. The effective height of the plateau scales as $c_0/3 \, \ln L$
with $c_0 > 1$ which indicates a logarithmic increase of the entropy for
these intermediate times.

\section{Summary and conclusions}
\label{sec:summary}

We have studied a particular kind of quench, where the system remained
critical and was only modified locally. Starting from the ground state 
with a defect, we calculated the evolution of the reduced density matrix 
and the entanglement following from it. The main effect is an increase
of the entropy $S$ on a time scale $t \approx L$ which can be related to the
excitations near the Fermi energy with velocity $v_F=1$ in our units. Thus
the entanglement always increases at the early stages of the rearrangement
process. The criticality of the system is reflected in the logarithmic 
dependence of various quantities, in particular the height of the maximum
of $S$, on the size of the subsystem. For the case of a boundary defect,
where the mechanism at work leads to a plateau, we could also find the
scaling function describing its shape. The defect strength only changes
the parameters, which are closely related to those in equilibrium,
but not the qualitative features.
The long-time behaviour is characterized by a slow approach to the 
homogeneous ground state. For the times accessible, one finds a universal
law independent of the defect strength. Thus no anomalous exponents as in 
the X-ray problem enter. These appear for time-delayed correlation functions
or in the overlap between initial and time-dependent state, whereas here one
is following the evolving state directly.
\par
A particular feature of our situation is that the initial state is asymptotically
degenerate with the final ground state. This distinguishes it from the case
of global quenches, where the energy difference between both is extensive, and also
from the study in \cite{Skrov06}, where the evolution started from a randomly
chosen high-energy state and the main interest was in the thermalization.
On the level of the 
reduced density matrix, one can define an effective temperature, if $\rho$
has thermal form, which is always the case for free fermions, and if the 
$\varepsilon_k$, or at least the important ones, are proportional to the 
single-particle excitations of the Hamiltonian. This is the case for critical
systems where both have linear dispersion \cite{EisLegRa06} and also for
the global quench considered in Appendix B. In this sense, our subsystem 
thermalizes with an effective temperature $T_{eff}=\ln L/\pi L$ which vanishes
if one considers the whole system,  while the quench in Appendix B leads
to a finite value $T_{eff}= 1/2$.
\par
Altogether we have obtained a good overall picture of the phenomena, although
some aspects like the oscillations in $S$ and the influence of the filling 
have not been addressed. However, an analytical derivation of the asymptotic
time dependence would still be desirable.
 
\ack We would like to thank M. Cramer, J. Eisert, E. Jeckelmann, C. Kollath,
A. L\"auchli and K.D. Schotte for discussions.

\section*{Appendix A: Initial entanglement}

For a defect at the boundary of the subsystem, the entanglement entropy
has already been studied in \cite{Peschel05}. For a general location, an
expression can be given immediately if $t'=0$. Then one has two half-chains 
where segments at the ends are part of the subsystem. If their respective
lengths $l$ and $L-l$ are large, the conformal formulae \cite{CC04} 
lead to
\eq{
S= \frac {c}{6}  \bigl[\,\ln(l)+ \ln(L-l) \, \bigr]+ 2 k_1
\label{eqn:defect1}
}
where $c=1$ and $k_1 \approx 0.479$ for the hopping model. Thus $S$ is maximal 
when the defect is in the center of the subsystem, $l =L/2$.
\par
\begin{figure}[thb]
\center
\includegraphics[scale=0.4,angle=270]{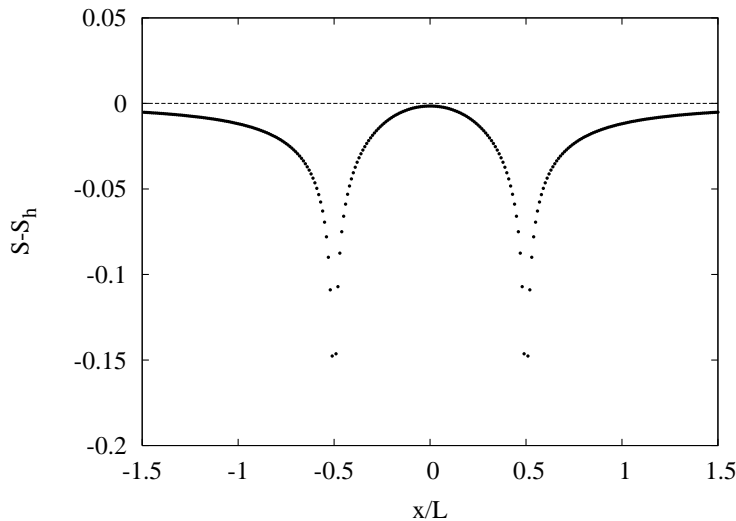}
\caption{Equilibrium entanglement entropy for a subsystem of size $L=200$
with a defect of $t'=0.5$ as a function of the defect position $x$ measured
from the center of the subsystem.}
\label{fig:ent_defectpos}
\end{figure}

Numerical calculations show that this law also holds for arbitrary defect
strengths $t' \leq 1$ but with a constant $c'$ depending on $t'$. Measuring the
position of the defect from the center of the subsystem, $l=L/2+x$, one can
write it in the form
\eq{
S= \bigl[\, \frac {1}{3} \ln L + k \,\bigr] + \frac {c'}{3} \ln \,\bigl[\,1 - 
(\frac {2x}{L})^2\;
\bigr], 
\label{eqn:defect2}
}
The term in the bracket with $k \approx 0.726$ represents the entropy of the 
homogeneous system which is reduced by the scaling function, except for a 
central defect. The coefficient $c'$ is related to the quantity $c_{eff}$ which 
appears for a boundary defect \cite{Peschel05} via $c'=1-c_{eff}$. If the defect 
is outside the subsystem, the same formula holds with $2x/L$ replaced by $L/2x$.
Then $S$ is unaffected, if the defect is far away from the subsystem, as
one expects. The situation is illustrated in Figure \ref{fig:ent_defectpos}
for the case $t'=0.5$ and served as a guidance for the analysis of the 
non-equilibrium results in Section \ref{sec:boundary}.
%
%
\section*{Appendix B: Homogeneous quench}

\begin{figure}[thb]
\center
\includegraphics[scale=0.4,angle=270]{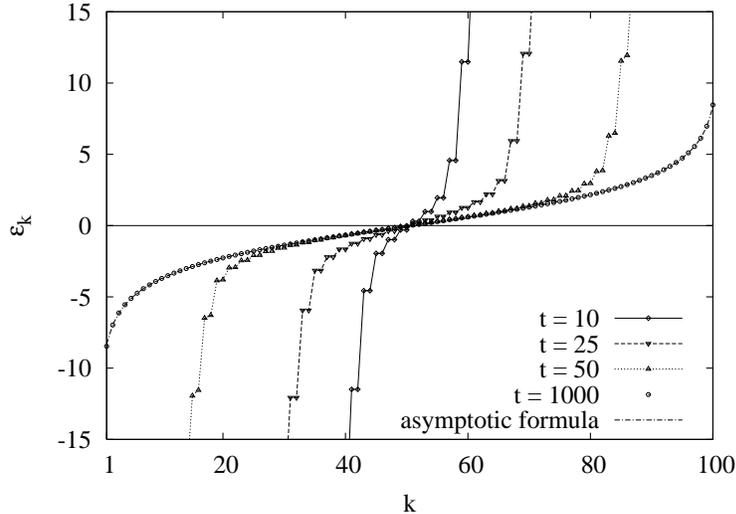}
\caption{Time evolution of the single-particle spectrum for a subsystem
of $L=100$ sites in case of a global quench starting with a fully dimerized
initial state.}
\label{fig:globalquench}
\end{figure}

It is instructive to compare local and global quenches at the level
of the density-matrix spectra. A simple global quench in the hopping
model is obtained if one starts with a fully dimerized chain in its ground
state which is then made homogeneous. Thus initially $t_{2n}= 1$ and 
$t_{2n+1}= 0$ and only pairs of sites are coupled. For half filling, 
this leads to a correlation matrix ${\bf{C}}(0)$ which consists of ($2 \times 2$)
blocks along the diagonal with all elements equal to one-half. Explicitly,
\eq{
C_{mn}(0) = \frac {1}{2}  \bigl[\delta_{m,n} + \frac {1}{2}(\delta_{n,m+1}+
\delta_{n,m-1}) +  \frac {(-1)^m}{2} (\delta_{n,m+1}-\delta_{n,m-1})\bigr]
\label{eqn:global1}
}
Due to the translational invariance (up to the alternating factors) this
leads to the simple result
\eq{
C_{mn}(t) = \frac {1}{2}  \bigl[\delta_{m,n} + \frac {1}{2}(\delta_{n,m+1}+
\delta_{n,m-1}) +  e^{-i \frac {\pi}{2}(m+n)} \frac {i(m-n)}{2t}J_{m-n}(2t) \bigr] 
\label{eqn:global2}
}
which involves only single Bessel functions. The resulting $\varepsilon_k$
are shown in Fig. \ref{fig:globalquench} for $L=100$ and several times.
%
\par
One sees that the dispersion
is linear near zero with a slope which decreases in time. This makes 
the number of low-lying levels larger and is responsible for the initial 
increase of the entanglement entropy. For times exceeding $L/2$, however, 
an asymptotic curve is approached and $S$ saturates. The asymptotic form of 
the $\varepsilon_k$ follows from the first two terms in (\ref{eqn:global2}) 
which describe a homogeneous tridiagonal matrix with eigenvalues
\eq{
\zeta_k(\infty)= \frac {1}{2} (1+\cos(p_k)), \;\;\; p_k =\frac {\pi\,k}{L+1},
 \;\;\; k=1,2...L  
\label{eqn:global3}
}
where the $p_k$ are the allowed momenta for an open chain. This gives  
\eq{
\varepsilon_k(\infty)= 2 \ln \tan(p_k/2)
\label{eqn:global4}
}
The spacing of the $p_k$ proportional to $1/L$ then leads to an asymptotic value
$S \sim L$ if one converts the sums for $S$ into integrals. The explicit result
is $S = L(2\ln2-1)$ and was found also in \cite{CC05} for a similar quench in the 
transverse Ising model. These results illustrate the strong influence of a
global quench on the form of the spectrum and the level spacing. For a local 
quench, Figure \ref{fig:epsilon_t} shows that the effects are much smaller 
and largely confined to intermediate times. 

\end{document}